\documentclass[11pt, a4paper]{article}   
\pdfoutput=4
\usepackage{jcappub} 
\usepackage{hyperref}
\usepackage{amsmath,mathtools}
\usepackage{verbatim}
\usepackage{enumerate}
\usepackage{graphicx}
\usepackage{xcolor}
\usepackage[utf8]{inputenc}
\usepackage[percent]{overpic}
\usepackage{multirow}
\usepackage{float}
\usepackage{booktabs}
\usepackage{ulem}
\usepackage[titletoc,title]{appendix}
\graphicspath{{newres-sbg/}}
\usepackage{comment,color}

\newcommand{\andreac}[1]{{\color{black}#1}}
\newcommand{\andrev}[1]{{\color{black}#1}}

\begin{document}
\title{IceCube Neutrinos from Hadronically Powered Gamma-Ray Galaxies}
\author{Andrea~Palladino, Anatoli~Fedynitch, Rasmus~W.~Rasmussen, Andrew~M.~Taylor}
\affiliation{DESY, Platanenallee 6, 15738 Zeuthen, Germany}
\date{\today}

\abstract{In this work we use a multi-messenger approach to determine if the high energy diffuse neutrino flux observed by the IceCube Observatory can originate from $\gamma$-ray sources powered by Cosmic Rays interactions with gas. Typical representatives of such sources are Starburst and Ultra-Luminous Infrared Galaxies. Using the three most recent calculations of the non-blazar contribution to the extragalactic $\gamma$-ray background measured by the Fermi-LAT collaboration, we find that a hard power-law spectrum with spectral index $\alpha \leq 2.12$ is compatible with all the estimations for the allowed contribution from non-blazar sources, within 1$\sigma$. Using such a spectrum we are able to interpret the IceCube results, showing that various classes of hadronically powered $\gamma$-ray galaxies can provide the dominant contribution to the astrophysical signal \andrev{above 100 TeV and about half of the contribution to the energy flux between 10-100 TeV}. With the addition of neutrinos from the Galactic plane, it is possible to saturate the IceCube signal \andrev{at high energy}. Our result \andrev{shows that these sources are still well motivated candidates.}} %reverses previous findings in which evidence was claimed against hadronic sources being the dominant source of IceCube neutrinos.}

\maketitle

\section{Introduction}
The IceCube Collaboration has detected a diffuse high-energy astrophysical neutrino flux \cite{Aartsen:2013jdh}. However, despite the six years since its detection, its origin remains unclear. Several candidate source populations have been proposed, such as blazars~\cite{bl1,bl2,bl3}, $\gamma$-Ray Bursts (GRBs)~\cite{grb1,Waxman:1997ti,grb4,grb5}, Star Forming Galaxies and Starburst Galaxies~\cite{Romero:2003tj,Loeb:2006tw,Liu:2013wia,Tamborra:2014xia}, dark matter decay~\cite{dm1,dm2} or Galactic sources, like Galactic center \cite{Razzaque:2013uoa, Ahlers:2013xia, Celli:2016uon}, Galactic plane \cite{Palladino:2016zoe,Pagliaroli:2016lgg,Pagliaroli:2017fse}, Galactic halo \cite{Taylor:2014hya}. Some models are strongly constrained by the absence of correlations between the directions of the high-energy neutrinos and known sources. This result is compatible with neutrino-bright sources being dim photon sources. For example, distant blazars that are not resolved by the Fermi satellite due to their low luminosity, could have a high fraction of interacting protons and thus be efficient neutrino sources \cite{Palladino:2018lov}. GRBs with chocked jets are another type of $\gamma$-ray dim and neutrino-bright sources \cite{Senno:2015tsn}. So far, only one neutrino source candidate has been tentatively identified, the blazar TXS 0506+056 \cite{IceCube:2018cha}, one of the brightest blazars located at a redshift $z=0.34$.

The multi-messenger approach can be applied in various ways, either on a ``per source'' basis where a single object is studied simultaneously in multiple wavelengths, in neutrinos and in gravitational waves. Alternatively, one can also study the entire populations of sources using diffuse (time and angular integrated) fluxes of neutrinos and photons, where the relative differences between observed energy spectra can constrain certain scenarios \cite{Murase:2013rfa}. The latter approach has been recently applied to study the contribution of Star Forming Galaxies (SFG), including Starburst and Star-Forming Active Galactic Nuclei, to the diffuse neutrino flux~\cite{Bechtol:2015uqb}. The study concluded that these sources can only contribute at most a few percent of the observed neutrinos, essentially excluding SFGs as the dominant sources of high-energy neutrinos in IceCube. \andrev{A similar conclusion is also reached in \cite{Murase:2015xka,Kistler:2015ywn}.}
Since the $\gamma$-ray radiation in SFGs predominantly originates from Cosmic Ray interactions with gas, the result may be generalized to the extent that  proton-proton ($pp$) interactions can not be the main mechanism of astrophysical neutrino production.

In astrophysical environments where pp interactions dominate cosmic ray cooling, the relation between the $\gamma$-ray and the neutrino emission is fixed. With approximately equal numbers of $\pi^+$, $\pi^-$ and $\pi^0$ being produced in high-energy collisions, the total energy budget of $\gamma$-rays is 2/3 of the total energy budget of neutrinos. 
If these ``hadronic sources'' power the entire diffuse neutrino flux, it is sufficient that only a subset of all galaxies significantly contributes, in line with the requirement for a high-energy cutoff of $E_\nu\sim$ few PeV. 
Such sources are likely to carry characteristics of Starburst galaxies (SBG) % or (ULIRGs). The superwinds and advectively dominated escape, etc.,..., cite, cite. 
such as enhanced stellar light and dust production, which results in infrared luminosities 10-100 times higher compared the more abundant normal galaxies \cite{Kewley:2001ng}. In reality, this heightened level of infrared galaxies is a continuum, with even higher values observed in what are classified as Ultra-Luminous Infrared Galaxies (ULIRGs), which are more luminous (infrared luminosity 100-1000 higher than normal Galaxies, see Fig.1 of \cite{Rojas-Bravo:2016val}) but less numerous. Fermi reveals an almost linear correlation between the infrared and the $\gamma$-ray luminosities \cite{Rojas-Bravo:2016val,Ackermann:2012vca}.
Therefore, ``infrared bright'' galaxies like Starburst and ULIRGs are good candidates of HAdronically powered $\gamma$-ray GalaxieS (HAGS). The neutrino and $\gamma$-ray spectrum from these objects is expected to be a power law, whose spectral index remains uncertain in the relevant energy range (above 50 GeV), since only a few members of the broad HAGS source class  (NGC253, M82) have been detected up to very high energies. We will therefore consider the spectral index as a free parameter in the present work, keeping in mind that the observation of NGC253 suggests an $\sim E^{-2.15 \pm 0.10}$ above 50 GeV, which is the energy region relevant for our purpose (see Fig.~\ref{fig:ngc}). 
%\textcolor{blue}{check the new plot in the supplementary material}
%\andrea{add discussion on the uncertainties on the spectral index for most of the HAGS. Only for 2 SBGs we know what is the high energy spectral index}

This work aims to re-evaluate the compatibility of the most recent neutrino observations by IceCube with an origin from HAGS. \andrev{We relax the hypothesis that the entire astrophysical neutrino flux is produced by a single source class, focusing first on the high energy part (above 100 TeV). Then }we apply a multi-messenger method and combine the diffuse extragalactic $\gamma$-ray background (EGB) observations by Fermi with the throughgoing muon energy flux measured by IceCube. \andrev{After that} we evaluate the agreement with other two IceCube dataset, where low energy events (down to $\sim$ TeV) are contained.

\section{Methods}
\label{sec:method}
We start with a comparison of the three IceCube data analyses, namely:
%We select three IceCube data analyses, in order to take into account information coming from different energy  \andreab{Again, this information is measleading. We do not care about what happens at low energy in our calculations. The contribution at lower energies comes out as a result. This part , marked in green, needs to be written from scratch}. \textcolor{green}{We use: 
\textit{i)} {\it the through-going muons (TGM) \cite{Aartsen:2016xlq,Aartsen:2017mau}} originating from muon neutrino and anti-neutrino interactions outside the detector. This selection contains tracks from the opposite (Northern) hemisphere and at a higher energy threshold of 200 TeV. \andreac{In \cite{Aartsen:2017mau} the throughgoing muon flux is fitted by using a power-law flux $\propto N\times E^{-\alpha}$, with $\alpha=2.2 \pm 0.1$ and $N=1.01^{+0.26}_{-0.23}$, where $N$ is the single flavor normalization at 100 TeV in units of $10^{-18} \ \rm GeV^{-1} cm^{-2} s^{-1} sr^{-1}$;}  \\
\textit{ii)} {\it the high-energy starting events (HESE) \cite{Aartsen:2013bka,Kopper:2017zzm}}, characterized by an interaction vertex contained in the fiducial detector volume. HESE contain shower and track like events 
%having a deposited energy larger than \anatoli{60} TeV \andrea{(@Anatoli: the energy threshold of HESE is 6000 photoelectrons, corresponding roughly to 30 TeV, not 60 TeV as stated by you. You can check this information on the science paper of 2013)} ,
mostly coming from the Southern hemisphere. \andreac{Concerning the HESE flux we refer to \cite{Aartsen:2017mau}, in which it has been found $\alpha=2.9 \pm 0.3$ and $N=2.5 \pm 0.8$ for this dataset, in the same units defined above};  \\
%, surrounded by a veto layer that improves the separation against Cosmic Ray muon background. The selection is sensitive to all neutrino flavors and is dominated by events from the Southern hemisphere, where the veto is effective. At low declinations a contamination with neutrinos of Galactic origin is expected \cite{Palladino:2016zoe} in addition to a residual atmospheric background (mainly cosmic muons) that partly comes from the inefficiency of the veto layer \cite{Palladino:2018evm}. The most recent update contains 5.7 (six) years of data with an energy threshold $\sim$60 TeV \cite{Kopper:2017zzm}; 
%, thus the sensitivity to a Galactic component is low at the respective declinations. Due to the high energy threshold of $\sim 200$ TeV, the contamination by conventional atmospheric neutrinos is low and only some prompt atmospheric neutrinos from decays of heavy flavor mesons can affect the measurement. TGM most likely represents the cleanest view on the high-energy astrophysical flux. 
\textit{iii)} {\it the 4-year cascade (CAS$_4$) \cite{Niederhausen:2017mjk}} sample contains neutrinos of all flavors interacting with a cascade topology, \textit{i.e.}\ mostly electron and tau neutrinos. The energy threshold is the lowest for this dataset, around $\sim$ TeV energies. \andreac{The flux associated to CAS$_4$ \cite{Aartsen:2017mau} is fitted by using $\alpha=2.48 \pm 0.08$ and $N=1.57^{+0.23}_{-0.22}$.}

We notice that above 200 TeV both throughgoing muons and HESE suggest a hard power law spectrum $d\Phi_\nu/dE \propto E^{-2.2 \pm 0.1}$. Since above this energy the contribution coming from the atmospheric background is small ($\sim 20\%$ using the signalness reported in Tab.4 of \cite{Aartsen:2016xlq}), we assume that this flux is representative of the true astrophysical signal. On the other hand one cannot neglect the information coming from the spectrum measured below 200 TeV; therefore, at the end of our multi-messenger analysis, we subsequently evaluate the compatibility between our findings with low energetic HESE and CAS$_4$ datasets.

In order to compute the diffuse neutrino and $\gamma$-ray fluxes we make use of the relation, in which the energy budget in neutrinos is 3/2 of that in $\gamma$-rays.  We assume for the density of the hadronic sources an evolution $\rho(z) = \rho_0(1+z)^{m}$, with $m=3.4$ for $z < 1$ and $m=-0.5$ up to $z=4$, like the star forming rate \cite{Yuksel:2008cu}.
The diffuse all-flavor neutrino flux $\phi_\nu$ is, therefore, related to the product of the local source density $\rho$ and the luminosity at 100 GeV $\mathcal{L}_{\text{100 GeV}}$, i.e.\ to the local emissivity: 
%This could be some sort of ``universal'' equation?! Values are random and would need to be replaced. We also need to decide if $\mathcal{L}$ refers to a specific energy or the integrated lumi. I think at 100 GeV is safer to avoid bolometric extrapolations. ANDREA: I am using in the paper the luminosity between 0.1 GeV and 3 TeV for NGC 253, as in the paper of Fermi and HESS. 
\begin{equation}
\small
\begin{split}
  \frac{d\Phi_\nu(E)}{dE} &= \int {\rm d}z~\frac{\rho(z)}{H(z)}~ \frac{d\phi_\nu}{dE}(E (1+z))\\
    &= \int {\rm d}z~\frac{\rho(z)}{H(z)}~\frac{3}{2}\frac{d\phi_\gamma}{dE}(E (1+z))  \\ 
    \approx \Phi_0 &\left (\frac{\rho_0}{\rho_0(\alpha)} \right ) \left ( \frac{\mathcal{L}_{\text{100 GeV}}}{\mathcal{L}_{\text{100 GeV}}^{\text{NGC253}}} \right )  \left(\frac{E}{\text{100 TeV}}\right)^{-\alpha^\star} e^{-E/E^\star_{\text{cut}}(\alpha)},
\end{split}
\end{equation}
\normalsize
where $\Phi_0$ is constant and equal to $3 \times 10^{-18} \ \rm GeV^{-1} \ cm^{-2} \ s^{-1} \ sr^{-1}$,
$\mathcal{L}_{\text{100 GeV}}^{\text{NGC 253}}=(5 \pm 2) \times 10^{39} \text{erg/s}$ is the $\gamma$-ray luminosity of NGC253 at 100 GeV \cite{Abdalla:2018nlz}, which we use here as a HAGS prototype. \andrev{We do not use any evolution of the $\gamma$-ray luminosity function, but instead assume a single prototype object for our calculation, which evolves only with an \lq\lq effective\rq\rq \ number density, such that the luminosity density evolves according to the Star Formation Rate.}
We determine the spectral-index dependent local ($z=0$) source density $\rho_0(\alpha)$ by fitting the astrophysical neutrino signal to the TGM spectrum in the range $0.1-1$ PeV.
More details on the calculation are reported in the supplementary material.
We assume throughout that the spectrum of neutrinos and $\gamma$-rays at the sources is described by a power-law with a spectral index $\alpha$ and an exponential cutoff energy  $E_{\text{cut}}$. The spectral index remains approximately the same at Earth while the energy cutoff is slightly different and we denote it with $E^\star_{\text{cut}}$. The values of $\rho(\alpha)$ and $E_{\text{cut}}^\star(\alpha)$, given in  Tab.~\ref{tab:phi0}, are discussed in the result section.

In the computation of the diffuse $\gamma$-ray flux, the interactions with the Extragalactic Background Light (EBL) have to be taken into account. The high-energy photons are absorbed and reprocessed through electromagnetic cascades, resulting in the re-appearance of this energy flux at lower energies. We use the same source spectrum as for the neutrinos, offset by 2/3 coming from energy budget considerations (see Eq.\ (\ref{eq:energy})). The transport of gamma rays through the intergalactic medium yields two components at the observer $\phi_\gamma = \phi_\gamma^{\text{dir}} + \phi_\gamma^{\text{casc}}$. The direct component here arrives from the source population, and is attenuated at
high energies by the EBL. This attenuation feeds electromagnetic cascades, giving
rise to the cascade component at energies below a few hundred GeV. We use \cite{Gilmore:2011ks} for the EBL model and the method given in \cite{Berezinsky:2016feh} to compute the electromagnetic cascade during the propagation.

The resulting propagated $\gamma$-ray flux has to be compared with the non blazar component of the extragalactic gamma-ray background (EGB). This residual component of the EGB  
%The extragalactic $\gamma$-ray background (EGB) is the remaining flux of $\gamma$-rays after the subtraction of individually sources  
was estimated by the Fermi collaboration \cite{TheFermi-LAT:2015ykq}. 
The remaining (non-blazar) fraction, is thought to be shared by all the other $\gamma$-ray emitters, such as normal and starburst galaxies and misaligned blazars. 
The share between the blazars and the non-blazar contribution varies between the different analyses, and are accompanied by large errors: the Fermi collaboration identifies the contribution from blazars to the EGB above 50 GeV as $86\% ^{+16}_{-14}\%$ \cite{TheFermi-LAT:2015ykq}; Lisanti et al. \cite{Lisanti:2016jub} as $68\%^{+9\%}_{-8\%}$ and Zechlin et al. \cite{Zechlin:2016pme} as $81\%^{+52\%}_{-19\%}$.
\andrev{Complementary to this measurement, the contribution from HAGS to the extragalactic $\gamma$-ray background has been discussed in \cite{Thompson:2006qd,Makiya:2010zt,Lacki:2012si}.}
The cumulative $\gamma$-ray flux at Earth $\phi_{\gamma}(\alpha)$ non-trivially depends on the spectral index of the source prototype since the share between the direct $\phi_{\rm dir}(\alpha)$ and the cascade $\phi_{\rm casc}(\alpha)$ components change. We evaluate the compatibility of HAGS with the non-blazar contribution by comparing the $\gamma$-ray flux integrated above 50 GeV with the integral of the EGB flux in the same energy range.

%the parameter $\mathcal{F}(\alpha)$: 
% \begin{equation}
% \mathcal{F}(\alpha)=\frac{\int_{50 \text{ GeV}}^\infty \frac{{\rm d} \phi_\gamma(\alpha)}{{\rm d}E} {\rm d}E }{\int_{50 \text{ GeV}}^\infty \frac{{\rm d} \phi_{\text{EGB}}}{{\rm d}E} {\rm d}E},
% \end{equation}
% which denotes the fractional contribution from ``hadronically powered $\gamma$-ray galaxies'' to the EGB as a function of the spectral index.

\begin{figure}[t]
    \centering
    \includegraphics[scale=0.4]{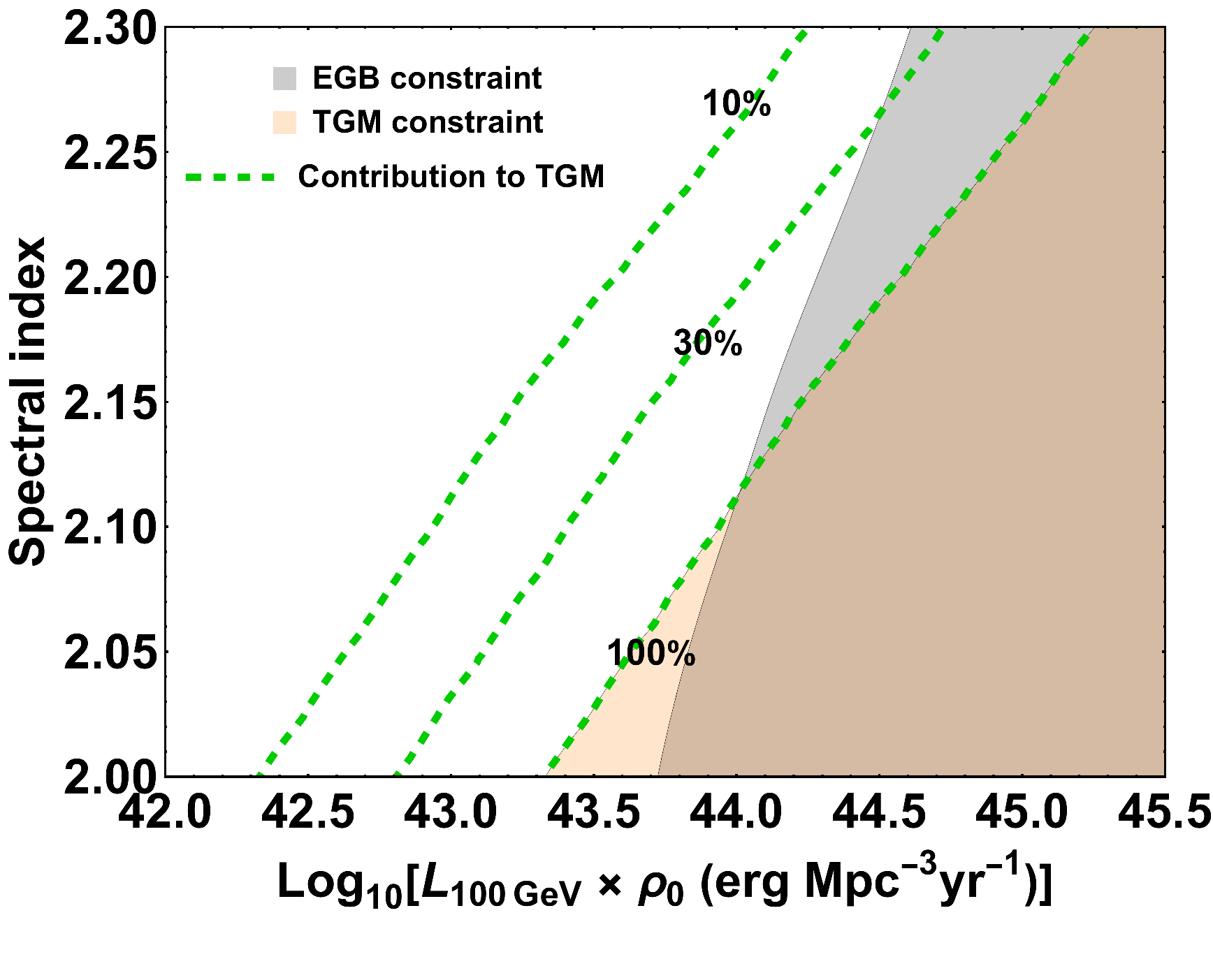}
    \caption{\textit{Summary of our main results. The grey band represents the region excluded taking into account the non blazar contribution to the EGB calculated by Fermi \cite{TheFermi-LAT:2015ykq}. The orange band represents the region excluded by integrating the throughgoing muon flux multiplied by energy between 200 TeV and 5 PeV, requiring that the energy budget predicted by our model is below the one measured by throughgoing muons. %plus the HESE number of events plus the shape of CAS$_4$ at low energy. 
    The green lines denote different contributions to the TGM energy flux.}}
    \label{fig:contour}
\end{figure}

\section{Results}
\label{sec:resultsa} We check the compatibility of our model with the TGM energy flux by scanning over the source spectral index $\alpha$ and by performing fits of the local density $\rho_0(\alpha)$ of HAGS that maximize the throughgoing muon energy flux.  We remark that the throughgoing muon sample is less contaminated by atmospheric muons and the 200 TeV energy threshold reduces the contribution of atmospheric neutrinos. Moreover no contamination from the Galactic plane is expected. 

In Fig.\ref{fig:contour} we report our result for different spectral indices and different luminosity densities. The bands represent forbidden regions at 1$\sigma$. The gray region indicates the exclusion of the model by its contribution to the most pessimistic non-blazar estimation of the EGB. %forbidden region bands (gray region), obtained by comparing the calculated EGB contribution with the most pessimistic non-blazar EGB estimation, are indicated. Likewise, the 1$\sigma$ forbidden region band, comparing the model integrated neutrino flux multiplied by energy 
The orange band limits the allowed neutrino energy flux integrated between
200 TeV and 5 PeV with respect to what is maximally allowed by the TGM measurements. The dashed green lines represent partial contributions to the TGM energy flux. The result is valid for generalized $pp$ sources, since the luminosity density is given by the product between the source density at redshift $z=0$ and the luminosity of the sources at 100 GeV. More details are reported in Tab.~\ref{tab:phi0} of the supplementary material. For harder spectrum and the two HAGS prototypes, the starburst galaxy NGC253 and the ULIRG Arp220 (with a luminosity 100-150 times higher than NGC \cite{Rojas-Bravo:2016val}), we find the local densities (given in Tab.\ref{tab:phi0}) to be compatible with the expectation for such sources. \andrev{Concerning Arp220, we remark that although this object is expected to be close to the calorimetric regime and ideal for neutrino production \cite{Yoast-Hull:2015iea,Wang:2016vue}, present measurements indicate a spectrum softer than $E^{-2}$ \cite{Peng:2016nsx}. In addition to this, there are also works discussing the possibility that this object is contamined by AGN activity \cite{Tamborra:2014xia,Yoast-Hull:2017wwl,Yoast-Hull:2019vro}, rather than a pure Starburst Galaxy.}
%In Tab.\ref{tab:phi0} we report for different spectral indices the normalization of the spectrum at 100 TeV $\phi_0$, the energy cutoff at Earth $E_{\text{cut}^\star(\alpha)}$ and the parameter $\rho_0(\alpha)$, 
%Given two HAGS prototypes, the Starburst Galaxy NGC253 and the ULIRG Arp220 (with a luminosity 100-150 times higher than NGC 253 \cite{Rojas-Bravo:2016val}),
%, Due to its distance,  has not been detected in the TeV band. Even Fermi observations are challenging \cite{Peng:2016nsx}, despite a $\gamma$-ray luminosity $\sim 150$ times larger than NGC 253. 
%for the harder spectra we find the local densities (given in Tab.\ref{tab:phi0}) to be compatible with the expectation for such sources \cite{Gruppioni:2013jna}. 

As demonstrated in the Supplementary Material, normal galaxies contribute very little to the diffuse $\gamma$-ray and the neutrino fluxes due to their soft spectral indices and low luminosities. Our result places an upper limit on the source spectral index, with values of $\alpha > 2.3$ being excluded at 5$\sigma$, thus confirming the findings by \cite{Bechtol:2015uqb} who based their arguments on the soft single power-law fit to the HESE data available at that time. On the contrary, we find that a source spectral index $\alpha \leq 2.12$ is found to be compatible with the three estimations of the non blazar contribution within $1\sigma$. Moreover an index of $\alpha=2.12$ allows the saturation of the throughgoing muon energy flux. 
%The right columns of Tab.\ref{tab:phi0} summarize the contribution of HAGS to the EGB. The tension with the three estimates of the non-blazar contribution by different groups is computed assuming Gaussian errors. 
\andrev{The first constraint on the spectral index of the neutrino flux is reported in \cite{Murase:2013rfa}, where the authors find $\alpha < 2.1-2.2$. Similar results were also reached in \cite{Chang:2014hua,Chang:2014sua,Ando:2015bva}. Our result slightly improves the previous constraints, although the uncertainty in the redshift evolution of the sources is not included and this may have some impact on the spectral index constraint, since the source evolution affects the diffuse $\gamma$-ray flux associated to neutrinos and, as a consequence, the comparison with the EGB.}

We note here that this discussion on the contribution to the non-blazar of the EGB neglects the inevitable contribution from misaligned AGN. An estimation of the level of this component is in the range between 4\% and 40\% \cite{DiMauro:2013xta}, with a best fit value of 12\%. Taking this into account the remaining non AGN contribution would sit at level of 16\%. Even for such a reduced level we still find compatibility with the throughgoing muon energy flux using a hard spectrum ($\alpha=2$) (see Appendix, Tab.\ref{tab:phi0}). \andrev{For the sake of completeness, we point out that in \cite{Sudoh:2018ana} it is shown that a hard spectrum would be the best scenario to reconcile the IceCube neutrinos above 100 TeV in the context of star-forming galaxies modeling, although it is not possible to saturate the flux under this hypothesis. This work is based on new theoretical model to predict the luminosity and spectrum of gamma-ray and neutrino emission from star-forming galaxies. The authors find that an injected $E^{-2}$ spectrum is able to power $\sim 22\%$ of the HESE events and $\sim 50\%$ of the throughgoing muon energy flux. An additional contribution from AGN-Starburst, rather than pure Starburst Galaxies, may help to saturate the observed astrophysical neutrino flux \cite{Bechtol:2015uqb}. In addition to, it has been discussed in the literature the possibility that high energy neutrinos can be produced by Galaxy Clusters \cite{Murase:2013rfa,Fang:2017zjf} and by Radio Galaxies \cite{Tjus:2014dna,Hooper:2016gjy}.}

\begin{figure}[t]
\centering
\includegraphics[scale=0.55]{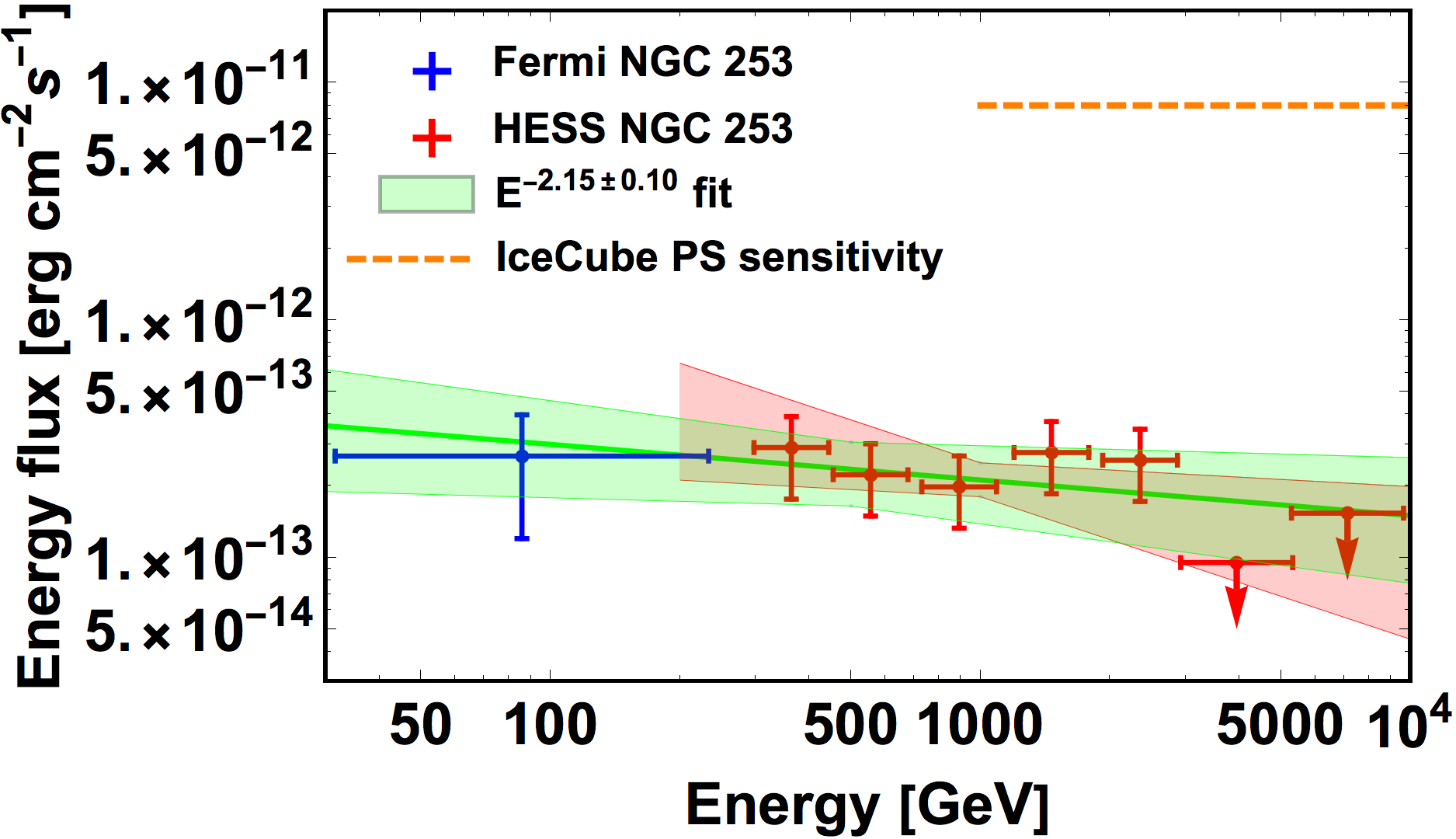}
\caption{\textit{The spectrum of the starburst galaxy NGC 253 from a combined analysis of Fermi (blue points) and HESS (red points and red band) \cite{Abdalla:2018nlz}. The green band represent the flux above 50 GeV. The observed data points are well fitted by an $E^{-2.15 \pm 0.10}$ spectrum, with an uncertainty of $\sim 25\%$ on the normalization.  %\textcolor{red}{ANDREA: Do you fit this spectrum? Where are the 30\% coming from? It was higher in the past, what happened?} \andrea{The uncertainties of 30\% reflects the uncertain of the throughgoing muon flux. The plot is different compared to the previous version, since we discussed it with Andrew, reaching the conclusion that for our purpose is not relevant what happens below 50 GeV.} 
The IceCube 7-year point source limit for a harder $E^{-2}$ spectrum and NGC253's declination $\delta \simeq -25^{\circ}$ is represented by the dashed line \cite{Aartsen:2016oji}.}}
\label{fig:ngc}
\end{figure}

\andrev{Concerning the $\gamma$-ray spectrum from NGC253, it was expected to be close to $\sim E^{-2.3}$ from theoretical arguments \cite{Yoast-Hull:2013qfa}.}
%On the other hand a too hard spectrum is disfavored by the observation of the Starburst Galaxy NGC 253. 
However the update measurements provided by Fermi and HESS \cite{Abdalla:2018nlz} suggest an $E^{-2.15 \pm 0.10}$ spectrum above 50~GeV, i.e.\ our energy region of interest (see Fig.\ref{fig:ngc}). \andrev{Therefore we rely on the observations, instead of considering theoretical predictions for the spectrum. Let us notice that such a hard spectrum can be explained from the theoretical point of view, assuming a small diffusion coefficient \cite{Lacki:2013nda}.} We re-fitted the data above 50 GeV, since we are particularly interested in the high-energy spectrum. The spectral indices obtained by  Fermi and HESS \cite{Abdalla:2018nlz} are $\alpha=2.09 \pm 0.07^{\text{stat}} \pm 0.05^{\text{sys}}$ above 60 MeV  and $\alpha=2.22 \pm 0.06^{\text{stat}}$ above 3 GeV, which both are compatible with our fit above 50 GeV.
Therefore as a baseline spectrum we choose $\alpha=2.12$, \textit{i)} being  the softest spectrum that is able to saturate the TGM energy flux, \textit{ii)} without producing any tension with the non blazar contribution, \textit{iii)} being compatible with the recent update on the observation result of NGC 253. We also remark that a spectral index $\alpha=2.12$ is perfectly consistent with the theoretical prediction presented almost 15 years ago in \cite{Loeb:2006tw}, in which a spectral index $\alpha= 2.15 \pm 0.10$ was expected for this kind of objects.

Moreover we also checked the compatibility of the local density resulting from our paper with the measured one. The infrared luminosity of NGC253 corresponds to $2 \times 10^{10} L_\odot$ (see Fig.~1 of \cite{Rojas-Bravo:2016val}), where $L_\odot$ is the luminosity of the Sun. For Starburst Galaxies with this infrared luminosity the local density is $1 \times 10^{-4} \rm \ Mpc^{-3}$ according to Fig.~10 of \cite{Gruppioni:2013jna}, i.e.\ 10 times smaller than that adopted in our benchmark model (see Tab.~\ref{tab:phi0}). However, one has to consider various elements: \textit{i)} this value depends on the spectral index; for example, the local density required for an $E^{-2}$ spectrum is $2 \times 10^{-4} \rm \ Mpc^{-3}$ (see Tab.~\ref{tab:phi0}); \textit{ii)} Starburst Galaxies are not the only sources contributing to the neutrino flux, since HAGS denote a larger class of objects, in which also Star Forming - AGN with Starburst characteristics are included. These objects are likely to be close to calorimetric, providing a hard neutrino spectrum. According to Tab.~1 and Tab.~2 in \cite{Tamborra:2014xia}, the local density of Star Forming - AGN (SB) is $ 1.5 \times 10^{-4} \rm \ Mpc^{-3}$; \textit{iii)} the relevant quantity is not the local density of the considered benchmark object, but the product between the local density and the $\gamma$-ray luminosity.
In our benchmark model, with $\rho_0 =10^{-3} \times \rm \ Mpc^{-3}$ and $L_\gamma^{0.1-100}= 6 \times 10^{39} \rm erg/s$ in the energy range 0.1-100 GeV (see Fig.~1 of \cite{Rojas-Bravo:2016val}), the required local luminosity density is
\begin{equation}
\label{eq:required_lumi}
\rho_0 L_\gamma^{0.1-100}= 2 \times 10^{44} \rm \ erg \ Mpc^{-3} \ yr^{-1},    
\end{equation}
to power the throughgoing muon flux. This value can be compared with Ref.~\cite{Murase:2016gly}, in which the luminosity function of Starburst galaxies has been explicitly included. From Fig.~3 of \cite{Murase:2016gly} relates a source density $\rho_0=10^{-5} \rm \ Mpc^{-3}$ to the luminosity $L_{\nu_\mu}^{10^4-10^7}=3 \times 10^{40}~{\rm erg~s}^{-1}$ in the energy range $10^{4}$ GeV and $10^7$ GeV. For what is discussed above, $L_\gamma \simeq 2/3 L_\nu$ in $pp$ interaction and considering the neutrino flavor equipartition coming from neutrino oscillations $L_\gamma=2 L_{\nu_\mu}$. To convert the luminosities computed in the different energy ranges, one obtains for an $E^{-2.2}$ spectrum $L_\gamma^{0.1-100} = 1/10 \ L_\gamma^{10^4-10^7}$. Therefore our result in Eq.~\ref{eq:required_lumi} is not in contradiction with \cite{Murase:2016gly}, demonstrating consistency of our simplified approach.%This value is compatible with the local density found for our benchmark scenario, that is reported in Tab.\ref{tab:phi0}.}%At multi-TeV energies the measured flux indicates a cutoff, reflecting a possible internal absorption of $\gamma$-rays \Andrew{NOT SURE THIS IS TRUE}.
%Soft spectral indices $\alpha > 2.3$ are excluded at 5$\sigma$ due to the EGB constraints, 

\begin{figure}[t]
    \centering
    \includegraphics[scale=0.41]{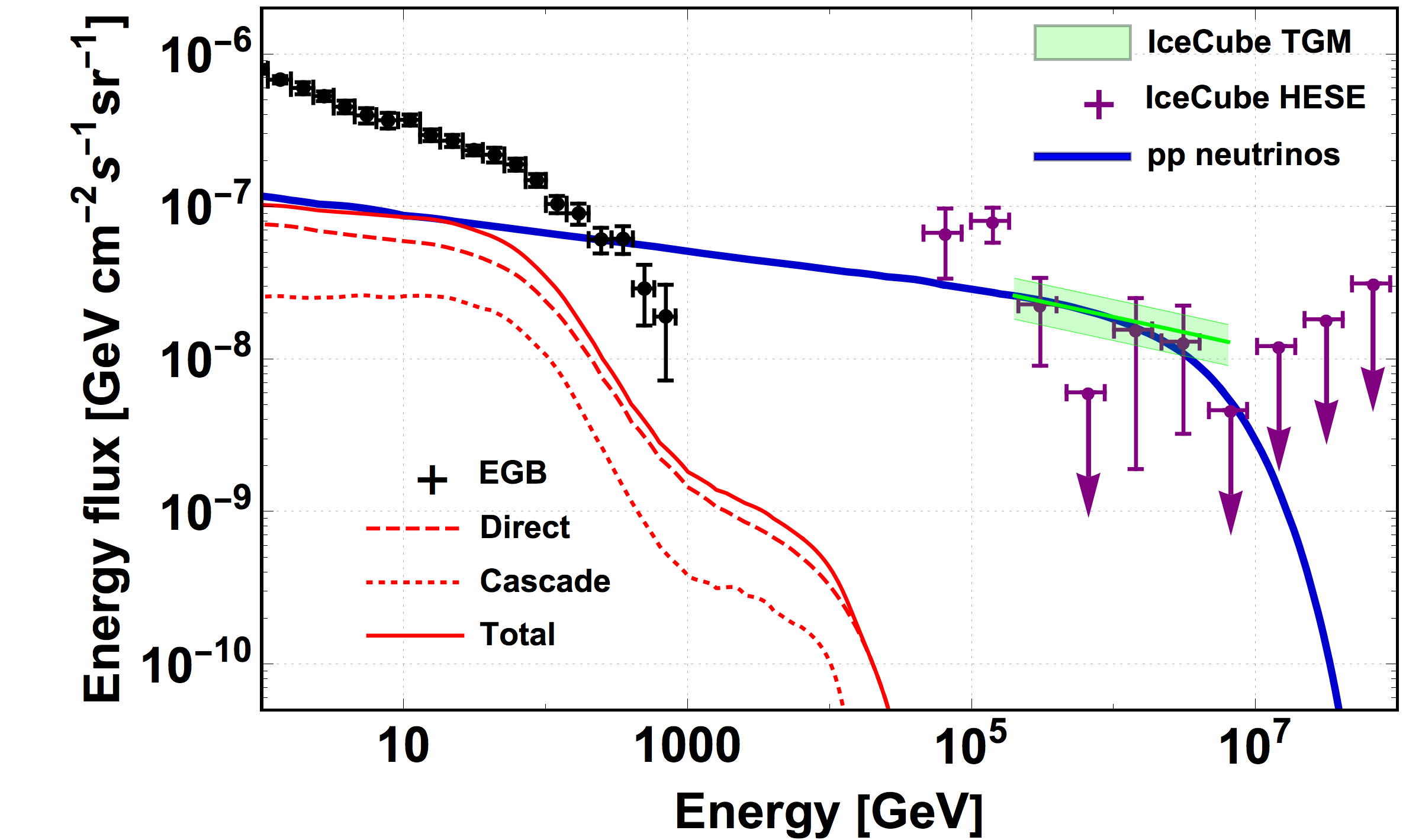}
    \caption{\textit{Diffuse extragalactic neutrino and $\gamma$-ray fluxes from a population of HAGS, using our baseline spectrum ($\alpha=2.12$) and 10 PeV as the energy cutoff at the source. The latest HESE data is indicated by the purple points \cite{Kopper:2017zzm} and the TGM energy flux by the green band \cite{Haack:2017dxi}. The black data points show the total Fermi EGB \cite{Ackermann:2014usa}. The $\gamma$-ray flux associated with HAGS is shown separately for the direct and the cascade fractions (red curves). The red curves are below the measurement since the integral flux above 50 GeV from HAGS can not exceed the non-blazar contribution accounting for a few tens of \% of the entire EGB.}}
    \label{fig:spec}
\end{figure}

Figure \ref{fig:spec} confirms that a neutrino spectrum of the form $dN/dE_{\nu}\propto E_{\nu}^{-2.12}$ with a source energy cutoff at 10 PeV is well in agreement with the TGM measurement and with the high-energy part of HESE.
%\andreab{I would not write this here but before, when we compare the three dataset. HESE and throughgoing muons are in agreement at high energy, therefore it is obvious that if we are in agreement with TGM we are also in agreement with HESE at high energy}. 
Smaller cutoff values are generally found to yield a better compatibility with the EGB, gradually falling short in explaining the highest energy neutrinos. An energy cutoff of 10~PeV for neutrinos at the source corresponds to an average cutoff of 200~PeV for protons. According to \cite{Romero:2003tj}, SBGs may be capable of accelerating protons to such high energies, supporting the compatibility of HAGS with the multi-messenger observations. \andrev{Moreover the possibility to accelerate protons up to the 100 PeV scale has been discussed in various works, invoking cosmic-ray accelerations due to hypernovae \cite{Chakraborty:2015sta,Senno:2015tra}, strong magnetic fields in Arp220-like galaxies \cite{Murase:2013rfa}, Galaxy mergers \cite{Kashiyama:2014rza,Yuan:2017dle} and AGN winds \cite{Tamborra:2014xia,Wang:2016vbf,Lamastra:2017iyo,Liu:2017bjr}.  
%a more effective accelerator members of the HAGS class is discussed in \cite{Romero:2018mnb}. 
Complementary to these expectations}, recent measurements by the two Ultra-High Energy Cosmic Ray (UHECR) observatories see first indications for a directional correlation between the arrival directions of cosmic rays above 39~EeV and nearby SBG \cite{Aab:2018chp, Abbasi:2018tqo}. Should HAGS be the sources of these UHECR, the local abundance of sources capable of reaching 200 PeV should be rather high.

\subsection{Comparison with low energy events}
We check the compatibility of our result with the other IceCube dataset. Concerning the low energy measurements, the expected number of HESE events are reported in Table~\ref{tab:events}. Using the HESE effective areas \cite{Aartsen:2013jdh} and our baseline spectrum, we find that HAGS can account for 33 out of about 41 astrophysical signal events, obtained by subtracting the $\sim$ 41 expected background events from the 82 detected events in HESE \cite{Aartsen:2017mau}. 
%\andreab{Maybe it is a good idae to clarify again that we obtain this as a result, because HESE and CAS$_4$ are not included in the fitting procedure, i.e.\ the comparison between EGB and throughgoing muon measurements.}
While the event counts are well described by our model, the hard spectrum undershoots the second HESE bin, leaving space for other small contributions from other sources or background in this energy range.
%\andreab{I see a significant tension only with the second HESE data point. By the way  the systematic uncertainty is even not represented...}. 
A possible additional contribution may come from the Milky Way's galactic disc can be present among the events below 100 TeV from the Southern hemisphere. Galactic neutrinos are expected to give a contribution below $\sim 150$ TeV, reflecting the possible $\sim 3$ PeV knee of the primary proton spectrum \cite{Palladino:2016zoe,Palladino:2016xsy,Pagliaroli:2017fse}. The current estimates predict $\sim$1~neutrino/year in the HESE dataset (above 30 TeV), in line with the latest experimental limits \cite{Albert:2018vxw}. Together with the galactic component, our model  saturates the HESE signal event count to 94\% at the best fit. 

\andrev{%We also compare our result with the differential event distribution of the CAS$_4$ sample. 
%(see discussion in Appendix \ref{appendix:cas4}). %In Fig.\ref{fig:cascade} we show the result and we can see that 
%[AF: the following has to be checked with the convolution of our atm prediction with the effective area.] 
%For the baseline spectral index $\alpha=2.12$, our model underestimates four points in a row (see Fig.\ref{fig:cascade}) between 10 TeV and 100 TeV, staying compatible within $1\sigma$.  Even in this case, the flux within this energy region expected from our model contributes more than 50\%. Therefore, the contribution from HAGS is dominant also in this energy region. Another contribution from additional gamma-ray-faint sources may be required to fully saturate the IceCube measurements below 100 TeV.
We also compare our result with the flux measured using the CAS$_4$ sample. 
As mentioned at the beginning of Sec.~\ref{sec:method}, the single flavor flux suggested by the low energy data is $N=1.57^{+0.23}_{-0.22}$ (the parameter $N$ is defined at the beginning of Sec.~\ref{sec:method}) and the spectral index is $\alpha=2.48 \pm 0.08$. On the other hand from our benchmark model we obtain a harder spectrum, having $N=1$ and $\alpha=2.12$. We compare the two energy fluxes, performing the ratio between the following integrals:  
$$
\mathcal{R}_{10-100} \equiv \frac{\int_{10 ~\text{TeV}}^{100 \text{ TeV}} \frac{d\Phi_\nu}{dE} \ dE}{\int_{10 \text{ TeV}}^{100 \text{ TeV}} \frac{d\Phi_{\text{CAS4}}}{dE} \ dE}
$$
obtaining $\mathcal{R}=0.40^{+0.07}_{-0.09}$ in the energy range between 10~TeV and 100~TeV, i.e.\ the sensitive energy range for the CAS$_4$ dataset. This suggests that even below 100~TeV the contribution from HAGS can be relevant, accounting for about half of the energy flux. For the remaining 50\% of the energy flux, various scenarios are plausible, such as a contamination from atmospheric prompt neutrinos \cite{Mascaretti:2019uqn}, from neutrinos produced in the Galactic plane \cite{Palladino:2016zoe,Pagliaroli:2016lgg} and/or from a second astrophysical component. However no conclusive interpretation can be provided with the present measurements.}

\section{Conclusion}
\label{sec:conclusion}
We applied a multi-messenger approach to study the contribution of HAGS to IceCube's diffuse neutrino flux combining the constraints from current $\gamma$-ray and neutrino observations. HAGS are (typically) infrared bright sources with a hard $\gamma$-ray spectrum from proton-proton interactions, for which the relation to the expected neutrino flux is well defined. The strongest constraint comes from the non-blazar contribution to the extragalactic $\gamma$-ray background.

We find that a hard power-law spectrum with an index $\alpha \leq 2.12$ and an energy cutoff at 10~PeV at the source to be compatible with the currently available estimations of the non-blazar contribution. In particular, due to the large uncertainties of these estimates it is impossible to derive tighter constraints solely from $\gamma$-ray observations and exclude HAGS, which include Starburst Galaxies and ULIRGs, as the dominant sources of diffuse neutrinos. Moreover, this conclusion remains valid even when an estimation of the misaligned AGN are further removed from the extragalactic $\gamma$-ray background. Such a spectrum is also found to be consistent with the spectrum of the NGC253 (a prototype of HAGS) above 50 GeV.

\begin{table}[t]
    \centering
    \caption{\textit{Decomposition of the 5.7 years of the 82 HESE events \cite{Kopper:2017zzm} into different source components. The background events are given by IceCube in \cite{Aartsen:2017mau}, whereas the Galactic neutrinos are computed in \cite{Pagliaroli:2016lgg} and they are roughly 1 per year in the HESE dataset. Neutrinos from HAGS accounts for 33 of the 41 signal events, i.e.\ about 80\% of the astrophysical signal.}
    \label{tab:events}}%\anatoli{Eeehm... how much is our model contributing? There is the important info missing.} \andrea{remove normal galaxies. say into the text that among HAGS the contribution of NG is small etc.. Put in the supplementary material the normal galaxy plot} 
    \begin{tabular}{ccc}
        \hline  
        & Expected & Fraction \\
        \hline
        Atmospheric muons & $25.2 \pm 7.3$  & $32\% \pm 9\%$ \\
        Atmospheric neutrinos & $15.6^{+11.4}_{-3.9}$  & $20\%^{+14\%}_{-5\%}$ \\
        Extragal. neutrinos (HAGS) & $33 \pm 8$  & $41\% \pm 10\%$ \\
        Galactic neutrinos & $\leq$ 5.7    & $\leq 7\%$ \\
        Total &  $79.5^{+15.9}_{-11.7}$  &--  \\
        \hline
    \end{tabular}
\end{table}

%\anatoli{[AF: This looks repetitive to me at this stage. I need to re-read the entire paper after we know more from the CAS4 atmospheric comparison.]}
\andreac{Concerning neutrinos, we find that our model can fully power the throughgoing muon energy flux detected by IceCube and the high energy part of the HESE flux, contributing to 80\% of the total signal contained in the HESE dataset and to more than 50\% of the signal contained in the CAS$_4$ dataset. %We also evaluate the contribution of our baseline spectrum to the lower energies, evaluating the compatibility with CAS$_4$ events. We find that more than 50\% of the events between 10 TeV and 100 TeV can be powered by HAGS, although an additional source may be required to fully interpret the measurements in this energy range. 
In conclusion,  
%\Andrew{TBD- where do we use all three?} For neutrinos we used three different data sets %\andreab{WRONG!}; \textcolor{green}{the through-going muon analysis for the spectral index and normalization at the highest energies, the event counts from the 5.7 year HESE analysis, and the differential event distribution from the new 4 year cascade analysis at intermediate and low energies.} 
we do not find any contradiction between the hypothesis that the dominant contribution of the IceCube neutrino energy flux comes from HAGS, since they can provide the dominant contribution in the energy range between 10 TeV and 10 PeV, where the neutrino flux is measured nowadays.} %since they can fully power the throughgoing muon energy flux and contribute to the majority of the signal neutrinos contained in HESE.
%\andreab{I would just say that HAGS can be the dominant source of high energy neutrinos, without repeating again the full discussion} In fact, HAGS may be the dominant source, saturating the throughgoing muon flux and providing the 80\% of the astrophysical flux contained in the HESE dataset \Andrew{Why do we want to discuss this here?}. 
The remaining fraction can be attributed to neutrinos from the Milky Way's galactic plane \andreac{plus other possible source contributing below 100 TeV}. Due to the soft spectral index and low maximal energies, the contribution from an entire population of ``normal galaxies'' (like ours) is at the few \% level. 
%Obtaining neutrino counts at the lower edge of the observed one is \Andrew{a good message- not sure what's meant by this???}, since we can not exclude that a few of these neutrinos come from other less abundant sources, such as gamma-ray bursts or active galactic nuclei.  \andreab{This sentence is obscure to me. I do not understand what is the message.}

This result \andrev{shows that relaxing the hypothesis that one single source class produces the entire astrophysical neutrino flux (as in \cite{Bechtol:2015uqb}), it is still allowed for sources characterized by $pp$ interaction to provide the dominant contribution to the IceCube neutrinos above 100 TeV and to produce half of the observed energy flux between 10-100 TeV. Therefore this result shows that HAGS are still worthy to be investigated in the context of astrophysical neutrinos, in agreement with the findings of previous works \cite{Anchordoqui:2014yva,Bartos:2015xpa,Chang:2016ljk,Xiao:2016rvd,Murase:2016gly,Chakraborty:2016mvc}.} %reverses the conclusions drawn in previous studies, which found evidence against ``Star Forming Galaxies'' \cite{Bechtol:2015uqb} as the major source of IceCube neutrinos. 

Compared to the brightest $\gamma$-ray sources (like blazars) HAGS are comparatively dim steady emitters, with their detection as neutrino sources requiring neutrino detectors with at least an order of magnitude more sensitivity compared to IceCube's current sensitivity. In the $\gamma$-ray domain, the upcoming Cherenkov Telescope Array (CTA) will be able to discover more nearby HAGS, as well as extend the energy spectral energy range up to higher energies of those already detected. Being hadronic sources, HAGS are natural candidates for this neutrino emission. We find good compatibility of HAGS being the dominant neutrino source class.

{\bf Acknowledgments.} This project has received funding from the European Research Council (ERC) under the European Union’s Horizon 2020 research and innovation programme (Grant No. 646623).

\bibliographystyle{ieeetr}
\bibliography{bibliography.bib}

\begin{thebibliography}{10}

\bibitem{Aartsen:2013jdh}
M.~G. Aartsen {\em et~al.}, ``{Evidence for High-Energy Extraterrestrial
  Neutrinos at the IceCube Detector},'' {\em Science}, vol.~342, p.~1242856,
  2013.

\bibitem{bl1}
R.~J. Protheroe, ``{High-energy neutrinos from blazars},'' {\em ASP Conf.
  Ser.}, vol.~121, p.~585, 1997.

\bibitem{bl2}
W.~Essey, O.~E. Kalashev, A.~Kusenko, and J.~F. Beacom, ``{Secondary photons
  and neutrinos from cosmic rays produced by distant blazars},'' {\em Phys.
  Rev. Lett.}, vol.~104, p.~141102, 2010.

\bibitem{bl3}
K.~Murase, Y.~Inoue, and C.~D. Dermer, ``{Diffuse Neutrino Intensity from the
  Inner Jets of Active Galactic Nuclei: Impacts of External Photon Fields and
  the Blazar Sequence},'' {\em Phys. Rev.}, vol.~D90, no.~2, p.~023007, 2014.

\bibitem{grb1}
B.~Paczynski and G.~H. Xu, ``{Neutrino bursts from gamma-ray bursts},'' {\em
  Astrophys. J.}, vol.~427, pp.~708--713, 1994.

\bibitem{Waxman:1997ti}
E.~Waxman and J.~N. Bahcall, ``{High-energy neutrinos from cosmological
  gamma-ray burst fireballs},'' {\em Phys. Rev. Lett.}, vol.~78,
  pp.~2292--2295, 1997.

\bibitem{grb4}
S.~Huemmer, P.~Baerwald, and W.~Winter, ``{Neutrino Emission from Gamma-Ray
  Burst Fireballs, Revised},'' {\em Phys. Rev. Lett.}, vol.~108, p.~231101,
  2012.

\bibitem{grb5}
K.~Murase and K.~Ioka, ``{TeV–PeV Neutrinos from Low-Power Gamma-Ray Burst
  Jets inside Stars},'' {\em Phys. Rev. Lett.}, vol.~111, no.~12, p.~121102,
  2013.

\bibitem{Romero:2003tj}
G.~E. Romero and D.~F. Torres, ``{Signatures of hadronic cosmic rays in
  starbursts? High-energy photons and neutrinos from NGC 253},'' {\em
  Astrophys. J.}, vol.~586, pp.~L33--L36, 2003.

\bibitem{Loeb:2006tw}
A.~Loeb and E.~Waxman, ``{The Cumulative background of high energy neutrinos
  from starburst galaxies},'' {\em JCAP}, vol.~0605, p.~003, 2006.

\bibitem{Liu:2013wia}
R.-Y. Liu, X.-Y. Wang, S.~Inoue, R.~Crocker, and F.~Aharonian, ``{Diffuse PeV
  neutrinos from EeV cosmic ray sources: Semirelativistic hypernova remnants in
  star-forming galaxies},'' {\em Phys. Rev.}, vol.~D89, no.~8, p.~083004, 2014.

\bibitem{Tamborra:2014xia}
I.~Tamborra, S.~Ando, and K.~Murase, ``{Star-forming galaxies as the origin of
  diffuse high-energy backgrounds: Gamma-ray and neutrino connections, and
  implications for starburst history},'' {\em JCAP}, vol.~1409, p.~043, 2014.

\bibitem{dm1}
A.~Esmaili and P.~D. Serpico, ``{Are IceCube neutrinos unveiling PeV-scale
  decaying dark matter?},'' {\em JCAP}, vol.~1311, p.~054, 2013.

\bibitem{dm2}
M.~Chianese, G.~Miele, S.~Morisi, and E.~Vitagliano, ``{Low energy IceCube data
  and a possible Dark Matter related excess},'' {\em Phys. Lett.}, vol.~B757,
  pp.~251--256, 2016.

\bibitem{Razzaque:2013uoa}
S.~Razzaque, ``{The Galactic Center Origin of a Subset of IceCube Neutrino
  Events},'' {\em Phys. Rev.}, vol.~D88, p.~081302, 2013.

\bibitem{Ahlers:2013xia}
M.~Ahlers and K.~Murase, ``{Probing the Galactic Origin of the IceCube Excess
  with Gamma-Rays},'' {\em Phys. Rev.}, vol.~D90, no.~2, p.~023010, 2014.

\bibitem{Celli:2016uon}
S.~Celli, A.~Palladino, and F.~Vissani, ``{Neutrinos and $\gamma$-rays from the
  Galactic Center Region after H.E.S.S. multi-TeV measurements},'' {\em Eur.
  Phys. J.}, vol.~C77, no.~2, p.~66, 2017.

\bibitem{Palladino:2016zoe}
A.~Palladino and F.~Vissani, ``{Extragalactic plus Galactic model for IceCube
  neutrino events},'' {\em Astrophys. J.}, vol.~826, no.~2, p.~185, 2016.

\bibitem{Pagliaroli:2016lgg}
G.~Pagliaroli, C.~Evoli, and F.~L. Villante, ``{Expectations for high energy
  diffuse galactic neutrinos for different cosmic ray distributions},'' {\em
  JCAP}, vol.~1611, no.~11, p.~004, 2016.

\bibitem{Pagliaroli:2017fse}
G.~Pagliaroli and F.~L. Villante, ``{A multi-messenger study of the total
  galactic high-energy neutrino emission},'' {\em JCAP}, vol.~1808, no.~08,
  p.~035, 2018.

\bibitem{Taylor:2014hya}
A.~M. Taylor, S.~Gabici, and F.~Aharonian, ``{Galactic halo origin of the
  neutrinos detected by IceCube},'' {\em Phys. Rev.}, vol.~D89, no.~10,
  p.~103003, 2014.

\bibitem{Palladino:2018lov}
A.~Palladino, X.~Rodrigues, S.~Gao, and W.~Winter, ``{Interpretation of the
  diffuse astrophysical neutrino flux in terms of the blazar sequence},'' 2018.

\bibitem{Senno:2015tsn}
N.~Senno, K.~Murase, and P.~Meszaros, ``{Choked Jets and Low-Luminosity
  Gamma-Ray Bursts as Hidden Neutrino Sources},'' {\em Phys. Rev.}, vol.~D93,
  no.~8, p.~083003, 2016.

\bibitem{IceCube:2018cha}
M.~G. Aartsen {\em et~al.}, ``{Neutrino emission from the direction of the
  blazar TXS 0506+056 prior to the IceCube-170922A alert},'' {\em Science},
  vol.~361, no.~6398, pp.~147--151, 2018.

\bibitem{Murase:2013rfa}
K.~Murase, M.~Ahlers, and B.~C. Lacki, ``{Testing the Hadronuclear Origin of
  PeV Neutrinos Observed with IceCube},'' {\em Phys. Rev.}, vol.~D88, no.~12,
  p.~121301, 2013.

\bibitem{Bechtol:2015uqb}
K.~Bechtol, M.~Ahlers, M.~Di~Mauro, M.~Ajello, and J.~Vandenbroucke,
  ``{Evidence against star-forming galaxies as the dominant source of IceCube
  neutrinos},'' {\em Astrophys. J.}, vol.~836, no.~1, p.~47, 2017.

\bibitem{Murase:2015xka}
K.~Murase, D.~Guetta, and M.~Ahlers, ``{Hidden Cosmic-Ray Accelerators as an
  Origin of TeV-PeV Cosmic Neutrinos},'' {\em Phys. Rev. Lett.}, vol.~116,
  no.~7, p.~071101, 2016.

\bibitem{Kistler:2015ywn}
M.~D. Kistler, ``{Problems and Prospects from a Flood of Extragalactic TeV
  Neutrinos in IceCube},'' 2015.

\bibitem{Kewley:2001ng}
L.~J. Kewley, M.~A. Dopita, R.~S. Sutherland, C.~A. Heisler, and J.~Trevena,
  ``{Theoretical modeling of starburst galaxies},'' {\em Astrophys. J.},
  vol.~556, pp.~121--140, 2001.

\bibitem{Rojas-Bravo:2016val}
C.~Rojas-Bravo and M.~Araya, ``{Search for gamma-ray emission from star-forming
  galaxies with Fermi LAT},'' {\em Mon. Not. Roy. Astron. Soc.}, vol.~463,
  no.~1, pp.~1068--1073, 2016.

\bibitem{Ackermann:2012vca}
M.~Ackermann {\em et~al.}, ``{GeV Observations of Star-forming Galaxies with
  \textit{Fermi} LAT},'' {\em Astrophys. J.}, vol.~755, p.~164, 2012.

\bibitem{Aartsen:2016xlq}
M.~G. Aartsen {\em et~al.}, ``{Observation and Characterization of a Cosmic
  Muon Neutrino Flux from the Northern Hemisphere using six years of IceCube
  data},'' {\em Astrophys. J.}, vol.~833, no.~1, p.~3, 2016.

\bibitem{Aartsen:2017mau}
M.~G. Aartsen {\em et~al.}, ``{The IceCube Neutrino Observatory - Contributions
  to ICRC 2017 Part II: Properties of the Atmospheric and Astrophysical
  Neutrino Flux},'' 2017.

\bibitem{Aartsen:2013bka}
M.~G. Aartsen {\em et~al.}, ``{First observation of PeV-energy neutrinos with
  IceCube},'' {\em Phys. Rev. Lett.}, vol.~111, p.~021103, 2013.

\bibitem{Kopper:2017zzm}
C.~Kopper, ``{Observation of Astrophysical Neutrinos in Six Years of IceCube
  Data},'' {\em PoS}, vol.~ICRC2017, p.~981, 2018.

\bibitem{Niederhausen:2017mjk}
H.~M. Niederhausen and Y.~Xu, ``{High Energy Astrophysical Neutrino Flux
  Measurement Using Neutrino-induced Cascades Observed in 4 Years of IceCube
  Data},'' {\em PoS}, vol.~ICRC2017, p.~968, 2018.

\bibitem{Yuksel:2008cu}
H.~Yuksel, M.~D. Kistler, J.~F. Beacom, and A.~M. Hopkins, ``{Revealing the
  High-Redshift Star Formation Rate with Gamma-Ray Bursts},'' {\em Astrophys.
  J.}, vol.~683, pp.~L5--L8, 2008.

\bibitem{Abdalla:2018nlz}
H.~Abdalla {\em et~al.}, ``{The starburst galaxy NGC 253 revisited by H.E.S.S.
  and Fermi-LAT},'' {\em Submitted to: Astron. Astrophys.}, 2018.

\bibitem{Gilmore:2011ks}
R.~C. Gilmore, R.~S. Somerville, J.~R. Primack, and A.~Dominguez,
  ``{Semi-analytic modeling of the EBL and consequences for extragalactic
  gamma-ray spectra},'' {\em Mon. Not. Roy. Astron. Soc.}, vol.~422, p.~3189,
  2012.

\bibitem{Berezinsky:2016feh}
V.~Berezinsky and O.~Kalashev, ``{High energy electromagnetic cascades in
  extragalactic space: physics and features},'' {\em Phys. Rev.}, vol.~D94,
  no.~2, p.~023007, 2016.

\bibitem{TheFermi-LAT:2015ykq}
M.~Ackermann {\em et~al.}, ``{Resolving the Extragalactic $\gamma$-Ray
  Background above 50 GeV with the Fermi Large Area Telescope},'' {\em Phys.
  Rev. Lett.}, vol.~116, no.~15, p.~151105, 2016.

\bibitem{Lisanti:2016jub}
M.~Lisanti, S.~Mishra-Sharma, L.~Necib, and B.~R. Safdi, ``{Deciphering
  Contributions to the Extragalactic Gamma-Ray Background from 2 GeV to 2
  TeV},'' {\em Astrophys. J.}, vol.~832, no.~2, p.~117, 2016.

\bibitem{Zechlin:2016pme}
H.-S. Zechlin, A.~Cuoco, F.~Donato, N.~Fornengo, and M.~Regis, ``{Statistical
  Measurement of the Gamma-ray Source-count Distribution as a Function of
  Energy},'' {\em Astrophys. J.}, vol.~826, no.~2, p.~L31, 2016.

\bibitem{Thompson:2006qd}
T.~A. Thompson, E.~Quataert, and E.~Waxman, ``{The Starburst Contribution to
  the Extra-Galactic Gamma-Ray Background},'' {\em Astrophys. J.}, vol.~654,
  pp.~219--225, 2006.

\bibitem{Makiya:2010zt}
R.~Makiya, T.~Totani, and M.~A.~R. Kobayashi, ``{Contribution from Star-Forming
  Galaxies to the Cosmic Gamma-Ray Background Radiation},'' {\em Astrophys.
  J.}, vol.~728, p.~158, 2011.

\bibitem{Lacki:2012si}
B.~C. Lacki, S.~Horiuchi, and J.~F. Beacom, ``{The Star-Forming Galaxy
  Contribution to the Cosmic MeV and GeV Gamma-Ray Background},'' {\em
  Astrophys. J.}, vol.~786, p.~40, 2014.

\bibitem{Yoast-Hull:2015iea}
T.~M. Yoast-Hull, J.~S. Gallagher, and E.~G. Zweibel, ``{Cosmic rays,
  $\gamma$-rays, and neutrinos in the starburst nuclei of Arp 220},'' {\em Mon.
  Not. Roy. Astron. Soc.}, vol.~453, no.~1, pp.~222--228, 2015.

\bibitem{Wang:2016vue}
X.~Wang and B.~D. Fields, ``{Are Starburst Galaxies Proton Calorimeters?},''
  {\em Mon. Not. Roy. Astron. Soc.}, vol.~474, no.~3, pp.~4073--4088, 2018.

\bibitem{Peng:2016nsx}
F.-K. Peng, X.-Y. Wang, R.-Y. Liu, Q.-W. Tang, and J.-F. Wang, ``{First
  detection of GeV emission from an ultraluminous infrared galaxy: Arp 220 as
  seen with the Fermi Large Area Telescope},'' {\em Astrophys. J.}, vol.~821,
  no.~2, p.~L20, 2016.

\bibitem{Yoast-Hull:2017wwl}
T.~M. Yoast-Hull, J.~S. Gallagher, S.~Aalto, and E.~Varenius, ``{Gamma-Ray
  emission from Arp 220: indications of an active galactic nucleus},'' {\em
  Mon. Not. Roy. Astron. Soc.}, vol.~469, no.~1, pp.~L89--L93, 2017.

\bibitem{Yoast-Hull:2019vro}
T.~M. Yoast-Hull and N.~Murray, ``{Breaking the Radio - Gamma-Ray Connection in
  Arp 220},'' {\em Mon. Not. Roy. Astron. Soc.}, vol.~484, no.~3,
  pp.~3665--3680, 2019.

\bibitem{Chang:2014hua}
X.-C. Chang and X.-Y. Wang, ``{The diffuse gamma-ray flux associated with
  sub-PeV/PeV neutrinos from starburst galaxies},'' {\em Astrophys. J.},
  vol.~793, no.~2, p.~131, 2014.

\bibitem{Chang:2014sua}
X.-C. Chang, R.-Y. Liu, and X.-Y. Wang, ``{Star-forming galaxies as the origin
  of the IceCube PeV neutrinos},'' {\em Astrophys. J.}, vol.~805, no.~2, p.~95,
  2015.

\bibitem{Ando:2015bva}
S.~Ando, I.~Tamborra, and F.~Zandanel, ``{Tomographic Constraints on
  High-Energy Neutrinos of Hadronuclear Origin},'' {\em Phys. Rev. Lett.},
  vol.~115, no.~22, p.~221101, 2015.

\bibitem{DiMauro:2013xta}
M.~Di~Mauro, F.~Calore, F.~Donato, M.~Ajello, and L.~Latronico, ``{Diffuse
  $\gamma$-ray emission from misaligned active galactic nuclei},'' {\em
  Astrophys. J.}, vol.~780, p.~161, 2014.

\bibitem{Sudoh:2018ana}
T.~Sudoh, T.~Totani, and N.~Kawanaka, ``{High-energy gamma-ray and neutrino
  production in star-forming galaxies across cosmic time: Difficulties in
  explaining the IceCube data},'' {\em Publ. Astron. Soc. Jap.}, vol.~70,
  no.~3, pp.~Publications of the Astronomical Society of Japan, Volume 70,
  Issue 3, 1 June 2018, 49, https://doi.org/10.1093/pasj/psy039, 2018.

\bibitem{Fang:2017zjf}
K.~Fang and K.~Murase, ``{Linking High-Energy Cosmic Particles by Black Hole
  Jets Embedded in Large-Scale Structures},'' {\em Phys. Lett.}, vol.~14,
  p.~396, 2018.
\newblock [Nature Phys.14,no.4,396(2018)].

\bibitem{Tjus:2014dna}
J.~Becker~Tjus, B.~Eichmann, F.~Halzen, A.~Kheirandish, and S.~M. Saba,
  ``{High-energy neutrinos from radio galaxies},'' {\em Phys. Rev.}, vol.~D89,
  no.~12, p.~123005, 2014.

\bibitem{Hooper:2016gjy}
D.~Hooper, T.~Linden, and A.~Lopez, ``{Radio Galaxies Dominate the High-Energy
  Diffuse Gamma-Ray Background},'' {\em JCAP}, vol.~1608, no.~08, p.~019, 2016.

\bibitem{Yoast-Hull:2013qfa}
T.~M. Yoast-Hull, J.~S.~G. III, E.~G. Zweibel, and J.~E. Everett, ``{Active
  Galactic Nuclei, Neutrinos, and Interacting Cosmic Rays in NGC 253 and NGC
  1068},'' {\em Astrophys. J.}, vol.~780, p.~137, 2014.

\bibitem{Lacki:2013nda}
B.~C. Lacki, ``{Sturm und Drang: Supernova-Driven Turbulence, Magnetic Fields,
  and Cosmic Rays in the Chaotic Starburst Interstellar Medium},'' 2013.

\bibitem{Gruppioni:2013jna}
C.~Gruppioni {\em et~al.}, ``{The Herschel PEP/HerMES Luminosity Function. I:
  Probing the Evolution of PACS selected Galaxies to z~4},'' {\em Mon. Not.
  Roy. Astron. Soc.}, vol.~432, p.~23, 2013.

\bibitem{Murase:2016gly}
K.~Murase and E.~Waxman, ``{Constraining High-Energy Cosmic Neutrino Sources:
  Implications and Prospects},'' {\em Phys. Rev.}, vol.~D94, no.~10, p.~103006,
  2016.

\bibitem{Aartsen:2016oji}
M.~G. Aartsen {\em et~al.}, ``{All-sky Search for Time-integrated Neutrino
  Emission from Astrophysical Sources with 7 yr of IceCube Data},'' {\em
  Astrophys. J.}, vol.~835, no.~2, p.~151, 2017.

\bibitem{Chakraborty:2015sta}
S.~Chakraborty and I.~Izaguirre, ``{Diffuse neutrinos from extragalactic
  supernova remnants: Dominating the 100 TeV IceCube flux},'' {\em Phys.
  Lett.}, vol.~B745, pp.~35--39, 2015.

\bibitem{Senno:2015tra}
N.~Senno, P.~Mészáros, K.~Murase, P.~Baerwald, and M.~J. Rees,
  ``{Extragalactic star-forming galaxies with hypernovae and supernovae as
  high-energy neutrino and gamma-ray sources: the case of the 10 TeV neutrino
  data},'' {\em Astrophys. J.}, vol.~806, no.~1, p.~24, 2015.

\bibitem{Kashiyama:2014rza}
K.~Kashiyama and P.~Meszaros, ``{Galaxy Mergers as a Source of Cosmic Rays,
  Neutrinos, and Gamma Rays},'' {\em Astrophys. J.}, vol.~790, p.~L14, 2014.

\bibitem{Yuan:2017dle}
C.~Yuan, P.~Mészáros, K.~Murase, and D.~Jeong, ``{Cumulative Neutrino and
  Gamma-Ray Backgrounds from Halo and Galaxy Mergers},'' {\em Astrophys. J.},
  vol.~857, no.~1, p.~50, 2018.

\bibitem{Wang:2016vbf}
X.~Wang and A.~Loeb, ``{Cumulative neutrino background from quasar-driven
  outflows},'' {\em JCAP}, vol.~1612, no.~12, p.~012, 2016.

\bibitem{Lamastra:2017iyo}
A.~Lamastra, N.~Menci, F.~Fiore, L.~A. Antonelli, S.~Colafrancesco, D.~Guetta,
  and A.~Stamerra, ``{Extragalactic gamma-ray background from AGN winds and
  star-forming galaxies in cosmological galaxy formation models},'' {\em
  Astron. Astrophys.}, vol.~607, p.~A18, 2017.

\bibitem{Liu:2017bjr}
R.-Y. Liu, K.~Murase, S.~Inoue, C.~Ge, and X.-Y. Wang, ``{Can winds driven by
  active galactic nuclei account for the extragalactic gamma-ray and neutrino
  backgrounds?},'' {\em Astrophys. J.}, vol.~858, no.~1, p.~9, 2018.

\bibitem{Aab:2018chp}
A.~Aab {\em et~al.}, ``{An Indication of anisotropy in arrival directions of
  ultra-high-energy cosmic rays through comparison to the flux pattern of
  extragalactic gamma-ray sources},'' {\em Astrophys. J.}, vol.~853, no.~2,
  p.~L29, 2018.

\bibitem{Abbasi:2018tqo}
R.~U. Abbasi {\em et~al.}, ``{Search for correlations between arrival
  directions of ultrahigh-energy cosmic rays detected by the Telescope Array
  experiment and a flux pattern from nearby starburst galaxies},'' {\em
  Submitted to: Astrophys. J. Lett.}, 2018.

\bibitem{Palladino:2016xsy}
A.~Palladino, M.~Spurio, and F.~Vissani, ``{On the IceCube spectral anomaly},''
  {\em JCAP}, vol.~1612, no.~12, p.~045, 2016.

\bibitem{Albert:2018vxw}
A.~Albert {\em et~al.}, ``{Joint constraints on Galactic diffuse neutrino
  emission from ANTARES and IceCube},'' 2018.

\bibitem{Mascaretti:2019uqn}
C.~Mascaretti and F.~Vissani, ``{On the relevance of prompt neutrinos for the
  interpretation of the IceCube signals},'' 2019.

\bibitem{Haack:2017dxi}
C.~Haack and C.~Wiebusch, ``{A measurement of the diffuse astrophysical muon
  neutrino flux using eight years of IceCube data.},'' {\em PoS},
  vol.~ICRC2017, p.~1005, 2018.

\bibitem{Ackermann:2014usa}
M.~Ackermann {\em et~al.}, ``{The spectrum of isotropic diffuse gamma-ray
  emission between 100 MeV and 820 GeV},'' {\em Astrophys. J.}, vol.~799,
  p.~86, 2015.

\bibitem{Anchordoqui:2014yva}
L.~A. Anchordoqui, T.~C. Paul, L.~H.~M. da~Silva, D.~F. Torres, and B.~J.
  Vlcek, ``{What IceCube data tell us about neutrino emission from star-forming
  galaxies (so far)},'' {\em Phys. Rev.}, vol.~D89, no.~12, p.~127304, 2014.

\bibitem{Bartos:2015xpa}
I.~Bartos and S.~Marka, ``{Spectral Decline of PeV Neutrinos from Starburst
  Galaxies},'' 2015.

\bibitem{Chang:2016ljk}
X.-C. Chang, R.-Y. Liu, and X.-Y. Wang, ``{How far are the sources of IceCube
  neutrinos? Constraints from the diffuse TeV gamma-ray background},'' {\em
  Astrophys. J.}, vol.~825, p.~148, 2016.

\bibitem{Xiao:2016rvd}
D.~Xiao, P.~Mészáros, K.~Murase, and Z.-g. Dai, ``{Revisiting the
  Contributions of Supernova and Hypernova Remnants to the Diffuse High-energy
  Backgrounds: Constraints on Very High Redshift Injection},'' {\em Astrophys.
  J.}, vol.~826, no.~2, p.~133, 2016.

\bibitem{Chakraborty:2016mvc}
S.~Chakraborty and I.~Izaguirre, ``{Star-forming galaxies as the origin of
  IceCube neutrinos: Reconciliation with Fermi-LAT gamma rays},'' 2016.

\bibitem{Gaggero:2014xla}
D.~Gaggero, A.~Urbano, M.~Valli, and P.~Ullio, ``{Gamma-ray sky points to
  radial gradients in cosmic-ray transport},'' {\em Phys. Rev.}, vol.~D91,
  no.~8, p.~083012, 2015.

\end{thebibliography}

\pagenumbering{gobble}

\appendix
\section{Supplementary material}
The next sections are dedicated to more detailed calculations of \textit{i)} the diffuse neutrino flux, \textit{ii)} the expected number of events in the HESE and CAS$_4$ data sets and \textit{iii)} the contribution from Normal Galaxies.

\subsection{Computation of diffuse neutrino flux}
\label{sec:appendix}
For hadronic, {\it i.e.}\ proton-proton, interactions the $\gamma$-ray and the all-flavor neutrino spectrum obey the following energy budget relation:
\begin{equation}
\int_0^\infty \frac{d\phi_\gamma}{dE} E \ dE \approx \frac{2}{3} \int_0^\infty \frac{d\phi_\nu}{dE} E \ dE
\label{eq:energy}
\end{equation}
where $\frac{dN}{dE}$ is the differential flux at the source. 
%\textcolor{red}{ANDREA: The entire calculation is hard to read. Please go through this chapter and only leave the content that is indeed used in the calculation.}\andrea{Can you be more specific ? Since I wrote this appendix, I am highly biased !}
The diffuse neutrino flux expected from HAGS, we start from the $\gamma$-ray luminosity spectrum from a single source. We assume that it is a power law with a spectral index equal to $\alpha+2$, an energy cutoff $E_{\text{cut}}$ and we denote it as $\frac{d\mathcal{L}_\gamma}{dE}(E,E_{\text{cut}},\alpha)$. For each value of $\alpha$, the related neutrino spectrum is normalized in order to reproduce $L_\nu=3/2 \ L_\gamma^{\text{NGC 253}}$ in the energy range between 0.1 GeV and 3 TeV, since we take the Starburst Galaxy NGC 253 as a benchmark object in our calculation. This object has been recently observed by both Fermi-LAT and HESS \cite{Abdalla:2018nlz}.
%$$
%\frac{d\mathcal{L}_\nu}{dE}(E,\alpha) = N(\alpha) E^{-\alpha+2} \exp\left(-\frac{E}{E_{\text{cut}}} \right)
%$$
%where $N(\alpha) \equiv \left[ \int_{0.1 \ \text{GeV}}^{3000 \ \text{GeV}} E^{-\alpha+1}  \exp\left(-\frac{E}{E_{\text{cut}}} \right) \right]^{-1} $. 
%In order to compute the flux from a single source, we use as a prototype the luminosity associated to the Starburst Galaxy NGC 253, detected by both HESS and Fermi-LAT \cite{Abdalla:2018nlz}. 
Under the previous assumptions the neutrino flux from a single source is given by:
\begin{equation}
\frac{d\phi_\nu}{dE}(E,E_{\text{cut}},\alpha,z) = \int_0^z~{\rm d} z ~\frac{\frac{d\mathcal{L}_\nu}{dE}[E(1+z),E_{\text{cut}},\alpha]}{4 \pi  D_c(z)^2(1+z)^2 E^2} 
\end{equation}
where $d\mathcal{L}/dE$ is in units of energy/sec. We denote the neutrino flux from a single source with $\phi(E_\nu)$. 
%\textcolor{red}{ The formula above doesn't have the correct units. Why is there energy*redshift square in the denominator? Do you need this form?} 
%\andrea{why are the units wrong ? we have $\rm erg/s \times cm^{-2} \ sr^{-1} erg^{-2} = erg^{-1} s^{-1} cm^{-2} sr^{-1}$ that are the dimensions of the flux...}
%\textcolor{red}{Also zmax of 10 is excessive. If redshifts beyond 6 do not make numerical differences, keep it there.}\andrea{OK, I agree}
where $z_{\text{max}}=6$, $D_c(z)=D_H \times d(z)$ is the comoving distance, $D_H$ is the Hubble distance and $d(z)=\int_0^z d\tilde{z} \ h(\tilde{z})^{-1}$. In the previous equation the terms $h(z)=\sqrt{\Omega_\lambda+\Omega_m(1+z)^3}$ with $\Omega_\lambda=0.73$ and $\Omega_m=0.27$.

%\subsection{Cosmological evolution of the sources}

In order to obtain the cumulative neutrino flux we 
%use the same evolution in redshift of the Star Formation Rate used in \cite{Yuksel:2008cu}, characterized by three pieces of power laws. 
parameterize the evolution of the sources in redshift as in \cite{Yuksel:2008cu}:
\begin{equation}
\frac{{\rm d}N}{{\rm d}V}(z)= \rho_{0} \left[ \left(1+z\right)^{a \eta}+\left(\frac{1+z}{B}\right)^{b\eta}+\left(\frac{1+z}{C}\right)^{c\eta} \right]^{1/\eta} \, ,
\end{equation}
\normalsize
where $a=3.4$, $b=-0.3$, $c=-3.5$, $B=5000$, $C=9$, $\eta=-10$. The parameter $\rho_{0}$ denotes the local density of sources at redshift ($z=0$). In our case $\rho_0(\alpha)$ is obtained by fitting data and depends on the spectral index (values are given in Tab.\ref{tab:phi0}). %The parameter $\Phi_0$ is defined in the main text, in Eq.\ (\ref{eq:phi0}).
%We use the density of the sources in units of $\mbox{Mpc}^{-3}$, denoting it at redshift $z=0$ with $\rho(\alpha)$. 
%changing the normalization at redshift $z=0$. We indicate the normalization at $z=0$ with $\rho_0(\alpha)$ and it denotes the local density of sources that is required to power the astrophysical neutrino flux. This value is a function of the spectral index, since if the spectrum is soft a larger number of sources is required. %star formation rate multiplied by a normalization factor $\Phi_0$, that accounts for the fraction of Star Forming Galaxies that are Starburst Galaxies.
The diffuse neutrino flux is indicated using $\Phi(E_\nu)$ and it is obtained from:
\begin{equation}
\small
\frac{d\Phi_\nu}{dE_\nu}(E,E_{\text{cut}},\alpha) = \int_0^{z_{\text{max}} }\frac{{\rm d}\phi_\nu}{{\rm d}E} (E,E_{\text{cut}},\alpha,z) \ \frac{{\rm d}N}{{\rm d}V}(z,\alpha) \ \frac{{\rm d}V}{{\rm d}z}\ dz 
\label{eq:cumspec}
\end{equation}
where $dV/dz$ gives the relation between the comoving volume and the redshift as follows, $\frac{dV}{dz}= 4\pi D_H^3 \frac{d^2(z)}{h(z)}$. We have chosen $z_{\text{max}}=6$ in our calculations but we have checked that larger choices of $z_{\text{max}}$ do not produce any significant impact on the calculations.

\begin{table*}[t]
    \footnotesize
    \centering
    \caption{\textit{Summary of our main results. In the table the values of different parameters are reported, namely: the energy cutoff $E_\text{cut}^\star$ in units of PeV, the local source density $\rho_0(\alpha)$ in units of $\mbox{Mpc}^{-3}$, the contribution to the extragalactic $\gamma$-ray background as a function of the spectral index $\alpha$. The parameter $\rho_0(\alpha)$ denotes the local density of sources that is required to power the TGM energy flux. Here we show two different examples, assuming as a prototype source NGC 253 and ARP 220. In the last three columns on the right we write what is the tension between the contribution to the EGB expected from our model and the results of the three calculations discussed in the paper \cite{TheFermi-LAT:2015ykq,Lisanti:2016jub,Zechlin:2016pme}.}}\label{tab:phi0}
    \begin{tabular}{cccc|cccc}
        \hline
        &  & NGC 253 & ARP 220 & Contribution to & &Tension with & \\
        $\alpha$  & $E_\text{cut}^\star$ & $\rho_0(\alpha)$ & $\rho_0(\alpha)$ & total EGB & Fermi coll. \cite{TheFermi-LAT:2015ykq} & Lisanti et al. \cite{Lisanti:2016jub} & Zechlin et al. \cite{Zechlin:2016pme}  \\
        \hline
        1.9   &3.2  &$7.7 \times 10^{-5}$ &$5.1 \times 10^{-7}$ &6\% & no & no & no \\
        2.0  &4.1 & $2.3 \times 10^{-4}$ &$1.6 \times 10^{-6}$ &11\%& no & no & no \\
        2.1  &5.1 &$7.9 \times 10^{-4}$ & $5.3 \times 10^{-6}$&25\% & 0.8$\sigma$ & no & 0.3$\sigma$\\
        2.12   &5.3 & $1.0 \times 10^{-3}$ &$6.7 \times 10^{-6}$ &28\% & 1$\sigma$ & no & 0.5$\sigma$ \\
        2.2  &6.3 & $2.8 \times 10^{-3}$ &$1.8 \times 10^{-5}$ &50\% & 2.6$\sigma$ & 2.1$\sigma$ & 1.7$\sigma$ \\
        2.3 &7.8 & $1.1 \times 10^{-2}$ &$7.2 \times 10^{-5}$ &100\% & 6.9$\sigma$ &8.7$\sigma$ & 4.8$\sigma$ \\
        \hline
    \end{tabular}
    \normalsize
\end{table*}

\begin{figure*}[t]
\centering
\includegraphics[scale=0.55]{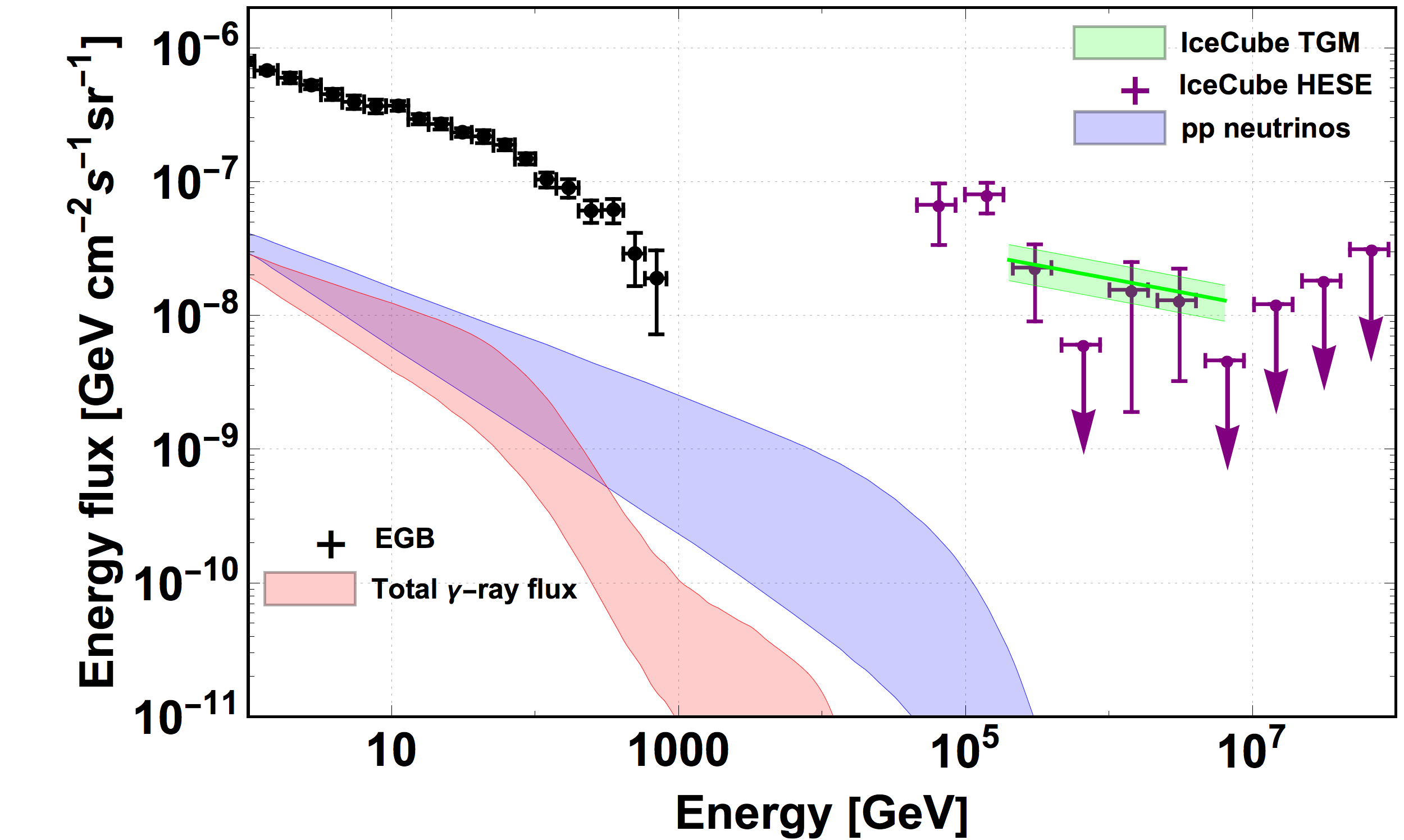}
\caption{\textit{$\gamma$-rays (red band) and neutrinos (blue band) produced by normal galaxies. The contribution to the EGB and to the IceCube flux is few $\%$, therefore normal Galaxies cannot be the dominant class of HAGS.}}
\label{fig:ng}
\end{figure*}

\subsection{Gamma-rays and neutrinos from Normal Galaxies}
Normal galaxies, such as ours, are more abundant but less luminous than HAGS. Therefore it might be important to estimate their contribution to the diffuse $\gamma$-ray and neutrino fluxes, using the same procedure as for HAGS. As a typical luminosity of normal galaxy we use that of the Milky Way $\sim 10^{39} \rm \ erg/s$ in the 0.1 GeV - 3 TeV energy range \cite{Rojas-Bravo:2016val}. The local density is estimated as $\rho_0=10^{-2.7} \ \rm Mpc^{-3}$ from {\it HERSCHEL} data (Figure 4 of \cite{Gruppioni:2013jna}).
The spectral index is varied in the range $\alpha \in [2.4-2.7]$, which includes the rather extreme KRA-$\gamma$ model \cite{Gaggero:2014xla} and that of galactic cosmic-rays. The resulting diffuse fluxes from normal galaxies are presented in Figure \ref{fig:ng} for the range of spectral indices. Even for the hardest spectral indices the contribution to the EGB and the neutrino flux is at level of few \%.
\end{document}